\documentclass[11pt]{article}

\usepackage{epsfig}
\usepackage{amsfonts,amssymb}

\newcommand{\sect}[1]{\setcounter{equation}{0}\section{#1}}

\newcommand{\al}{\ensuremath{\alpha}}

\newcommand{\Ga}{\ensuremath{\Gamma}}

\newcommand{\ep}{\ensuremath{\epsilon}}

\newcommand{\ka}{\ensuremath{\kappa}}

\newcommand{\La}{\ensuremath{\Lambda}}
\newcommand{\om}{\ensuremath{\omega}}
\newcommand{\Om}{\ensuremath{\Omega}}
\newcommand{\p}{\ensuremath{\phi}}

\renewcommand{\t}{\ensuremath{\tau}}
\renewcommand{\th}{\ensuremath{\theta}}

\renewcommand{\d}{\ensuremath{{\rm d}}}

\newcommand{\ra}{\ensuremath{\rightarrow}}
\newcommand{\del}{\ensuremath{\partial}}

\newcommand{\td} {\ensuremath{\tilde}}
\newcommand{\inv}{\ensuremath{^{-1}}}
\newcommand{\half}{\ensuremath{\frac{1}{2}}}

\newcommand{\be}{\begin{equation}}
\newcommand{\ee}{\end{equation}}

\newcommand{\ba}{\begin{eqnarray}}
\newcommand{\ea}{\end{eqnarray}}

\begin{document}

\bigskip
\hskip 4.8in\vbox{\baselineskip12pt
\hbox{hep-th/0302nnn} \hbox{SUSX-TH/03-003}}

\bigskip
\bigskip
\bigskip

\begin{center}
{\Large \bf Decay Modes of Intersecting Fluxbranes}\\
\end{center}

\bigskip
\bigskip
\bigskip

\centerline{\bf D. Brecher$^\natural$ and P. M. Saffin$^\flat$}

\bigskip
\bigskip
\bigskip

\centerline{\it $^\natural$Department of Physics and Astronomy}
\centerline{\it University of British Columbia}
\centerline{\it Vancouver, British Columbia V6T 1Z1, Canada}
\centerline{\small \tt brecher@physics.ubc.ca}

\centerline{$\phantom{and}$}

\centerline{\it $^\flat$Centre for Theoretical Physics}
\centerline{\it University of Sussex, Falmer, Brighton BN1 9QJ, U.K.}
\centerline{\small \tt P.M.Saffin@sussex.ac.uk}

\bigskip
\bigskip

\begin{abstract}
\vskip 2pt
Just as the single fluxbrane is quantum mechanically unstable to
the nucleation of a locally charged spherical brane, so intersecting fluxbranes are
unstable to various decay modes.  Each individual element of the
intersection can decay \emph{via} the nucleation of a spherical brane,
but uncharged spheres can also be nucleated in the region of
intersection.  For special values of the fluxes, however,
intersecting fluxbranes are supersymmetric, and so are expected to be
stable.  We explicitly consider the instanton describing the decay
modes of the two--element intersection (an F5-brane in the string theory
context), and show that in dimensions greater than four the action 
for the decay mode of the supersymmetric intersection diverges.
This observation allows us to show that stable intersecting
fluxbranes should also exist in type 0A string theory.
\end{abstract}

\newpage

\baselineskip=18pt
\setcounter{footnote}{0}

%%%%%%%%%%%%%%%%%%%%%%%%%%%%%%%%%%%%%%%%%%%%%%%%%%%%%%%%%%%%%%%%%%%%%%%%%%%%%%%

\sect{Introduction}

The four--dimensional Melvin universe~\cite{melvin:64} describes a
gravitating magnetic fluxtube which, in the context of Kaluza--Klein
theory~\cite{gibbons:88,gibbons:87}, can be obtained from five--dimensional
Minkowski space by identifying coordinates in a non--standard
way~\cite{dowker:9309,dowker:9312,dowker:9507,dowker:9512}.  By adding
six flat spectator dimensions, one obtains~\cite{dowker:9512,costa:0012} a
seven--dimensional fluxbrane (the F7-brane) which can be generated
\emph{via} a similar identification of points in eleven--dimensional
Minkowski space.  The F7-brane, and its intersecting cousins to be
considered below, are thus expected to be exact solutions of
M-theory, including all curvature corrections.  For the same reason,
string theory on the similar solution
with Neveu--Schwarz--Neveu--Schwarz (NS--NS) flux obtained from
ten--dimensional Minkowski space is exactly
solvable~\cite{russo:94,russo:95,tseytlin:95,russo:98}.

One can further generate spherical branes
immersed within the background of a single fluxbrane
\cite{dowker:9507,dowker:9512},
spherical D6-branes within the context of string theory, and supergravity
solutions~\cite{costa:01,emparan:0105,brecher:01} describing the dielectric
effect of Myers~\cite{myers:99}.  Of course, a spherical brane is
held only in metastable equilibrium by the ambient flux; it is
easy to see that the fluxbrane itself will be unstable to the
semi--classical nucleation of just such a spherical brane.  In fact, there is a second
decay mode~\cite{dowker:9507,dowker:9512}, corresponding to Witten's bubble of
nothing~\cite{witten:82}.

More explicitly, to obtain a flux $(D-4)$-brane from $D$ dimensional
flat space in terms of cylindrical polar coordinates, one
dimensionally reduces by identifying points separated by $2\pi R$
along the closed orbits of the Killing vector
\be
K = \frac{\del}{\del z} + B \frac{\del}{\del \p},
\ee

\noindent where the parameter $B$ becomes the magnetic field in the
reduced theory.  In other words, the identification of points is
\be
z \equiv z + 2 \pi n R, \qquad \p \equiv \p + 2 \pi n B R + 2 \pi
m \qquad n, m \in \mathbb{Z}.
\label{eqn:twisted}
\ee

\noindent These identifications are unchanged under $B \ra B +
n/R$, for some integer $n$, so physically inequivalent spacetimes
have~\cite{dowker:9507,dowker:9512}
\be
-\frac{1}{2R} \le B < \frac{1}{2R}.
\ee

\noindent In a theory with fermions, however, one must be more
careful~\cite{costa:0012}. Fermionic boundary conditions are
unchanged only under $B \ra B + 2n/R$, so we should consider the magnetic
field to lie in the range
\be
-\frac{1}{R} \le B < \frac{1}{R}.
\ee

\noindent This becomes clearer when one thinks about spin
structures~\cite{costa:0012}.  The original $D$--dimensional
space is topologically $\mathbb{M}^{D-1} \times S^1$, so there are
two distinct spin structures; under parallel transport around the
orbits of $K$, spinors pick up a phase
\be
+ e^{\pi R B \Ga} \qquad {\rm or} \qquad - e^{\pi R B \Ga},
\ee

\noindent where $\Ga$ is an element of the Lie algebra of ${\rm
Spin}(1,D-1)$ which generates rotations along $\p$ and has $\Ga^2 =
-1$.  As in the usual Kaluza--Klein scenario~\cite{witten:82}, we are
free to choose the overall sign.  For $B=0$, and in the string theory
context, the first choice
corresponds to the type IIA theory with supersymmetric boundary
conditions and the second corresponds to the non--supersymmetric type
0A theory~\cite{bergman:99}.  By continuity we make the same
assignments for non--zero $B$.

Since the twisted identifications (\ref{eqn:twisted}) break all
supersymmetry, there is no immediate obstruction to decay processes.
The two semi--classical decay modes of the
fluxbrane~\cite{dowker:9507,dowker:9512} are determined by choosing a spin
structure~\cite{costa:0012}.
In both cases~\cite{dowker:9507,dowker:9512}, the relevant instanton
is the Euclidean version of the $D$--dimensional Kerr black
hole found by Myers \& Perry~\cite{myers:86}.  We denote this as
Kerr$(\Omega)$, where $\Omega$ is the Euclidean ``rotation'' parameter.

To avoid a conical singularity in Kerr$(\Omega)$, the
Euclidean time coordinate must be periodic in the usual way.  But
one must also identify coordinates as in (\ref{eqn:twisted}), with 
$B=\Om$, where $\Omega$ lies in the range~\cite{dowker:9507,dowker:9512}
\be
-\frac{1}{R} < \Omega < \frac{1}{R}.
\ee

\noindent Reducing Kerr$(\Omega=B)$ along 
\be
\label{killing}
K = \frac{\del}{\del z} + B \frac{\del}{\del \p},
\ee

\noindent thus gives a $(D-1)$--dimensional solution with the same
asymptotic magnetic field, $B$, as reducing Kerr$(\Omega=B \mp 1/R)$ along 
\be
\label{killing'}
K' = K \pm \frac{1}{R} \frac{\del}{\del \p}.
\ee

\noindent The latter is the ``shifted instanton''
of~\cite{dowker:9507,dowker:9512}, and we use the upper (lower) sign
if $B$ is positive (negative).  As far as the bosonic fields are
concerned, one would thus conclude that a fluxbrane with magnetic field
$B$ has two decay modes, corresponding to the shifted and unshifted instanton.
However, due to the shift in $B$, the two
$(D-1)$--dimensional theories will have opposite spin structures.  To
determine which is which~\cite{dowker:9507}, note that under parallel transport around
integral curves at infinity of (\ref{killing}), spinors in
Kerr$(\Omega=B)$ pick up a phase $- \exp \left( \pi R B \Ga \right)$
so, in the string theory context, this corresponds to the 0A theory.  On the other hand, under
parallel transport around integral curves at infinity of (\ref{killing'}), spinors in
Kerr$(\Omega=B \mp 1/R)$ pick up a phase $- \exp \left( \pi R (B \mp
1/R) \Ga \right) = +\exp \left( \pi R B \Ga \right)$, which
corresponds to the IIA theory.

We thus have the following
picture~\cite{dowker:9507,dowker:9512,costa:0012}.  Reducing
Kerr$(\Omega=B)$ along $K$ gives the decay mode relevant to the
F7-brane in
the 0A theory; the fixed point set of $K$ is a nine--sphere so, upon
analytically continuing one of the ignorable angles, 
this describes a deformed version of Witten's
bubble of nothing~\cite{witten:82}.  Reducing instead Kerr$(\Omega=B
\mp 1/R)$ along $K'$ gives the decay mode relevant to the F7-brane in
the IIA theory; the fixed point set of $K'$ gives rise to an expanding
six--sphere upon analytic continuation.

As we will review below, intersecting fluxbranes can be constructed
\emph{via} a similar twisted reduction of flat Minkowski space; one
simply adds extra rotations to the Killing vector along which one
dimensionally reduces~\cite{dowker:9512}.  But here there is a
surprise, in that supersymmetry can be preserved for specific values of
the magnetic fields associated with each of the individual fluxbranes.  In the
ten--dimensional context of two intersecting F7-branes, this was first
pointed out in~\cite{gutperle:01} and a complete classification
of the ten--dimensional supersymmetric fluxbrane configurations was made
in~\cite{farrill:0110}.  Yet instantons describing the semi--classical
decay modes of intersecting fluxbranes certainly exist, having already
been discussed in~\cite{dowker:9512}; they are constructed from the
Euclideanized Myers--Perry black holes~\cite{myers:86} with more than one
plane of rotation\footnote{Upon a further analytic continuation of one
of the ignorable angles, and after dimensional reduction, these spaces
describe the evolution of intersecting fluxbranes after they decay
but, prior to dimensional reduction, they are interesting potential string theory
backgrounds in their own right~\cite{aharony:02}.}.

The question we want to address here is what happens to the instantons
describing these decay modes for the cases in which supersymmetry is
preserved.  Some mention of these issues has already appeared in~\cite{russo:01},
where string theory on intersecting NS--NS fluxbranes is shown to be
solvable, just as for the single fluxbrane.  As in that paper, one can
argue that the semi--classical amplitude for the
decay of a supersymmetric
solution must vanish, due to the presence of fermion zero--modes.
For the dual solution in type 0A however, there is no supersymmetry,
so stability does not arise from fermion zero modes. For this reason
we calculate the action of the instantons, showing that in the supersymmetric
case the action diverges, giving a vanishing semi--classical decay amplitude.

This note is organized as follows.  In the next section, we briefly
review the construction of the flux $(D-6)$-brane (two intersecting
flux $(D-4)$-branes),
before turning to consider the instantons describing the various decay
modes in section \ref{sec:instanton}.  We discuss spin structures and how they
determine the possible decay modes in section \ref{sec:spin} and compute
the action for the instanton in section \ref{sec:action}.  We conclude in
section \ref{sec:conclusion}.  We try to be dimension--independent
although, since we will ultimately be interested in string theory, we
will sometimes specialize to the ten--dimensional case.

%%%%%%%%%%%%%%%%%%%%%%%%%%%%%%%%%%%%%%%%%%%%%%%%%%%%%%%%%%%%%%%%%%%

\sect{Intersecting fluxbranes}
\label{sec:f5}

Although most of what we discuss here can be made more general, we
will concentrate on the case of two flux $(D-4)$-branes intersecting
over a flux $(D-6)$-brane.  To generate such a solution~\cite{dowker:9507,dowker:9512}, we start with
$D$--dimensional Minkowski space written as
\be
\label{eqn:11Dflat}
\d s^2_D = \d s^2 (\mathbb{M}^{D-5}) + \sum_{i=1,2} \left( \d \rho_i^2 +
\rho_i^2 \d \p_i^2 \right) + \d z^2,
\label{eqn:flat}
\ee

\noindent and, with $i=1,2$, make the identifications
\be
z \equiv z + 2 \pi n R, \qquad \p_i \equiv \p_i + 2 \pi n B_i R + 2 \pi
m_i \qquad n, m_i \in \mathbb{Z}.
\label{eqn:twotwists}
\ee

\noindent At least as far as the bosons are concerned, inequivalent
spacetimes are thus obtained for both
\be
-\frac{1}{2R} \le B_i < \frac{1}{2R}.
\ee

\noindent Geometrically, we dimensionally reduce along the closed
orbits of the Killing vector
\be
K = \frac{\del}{\del z} + B_1 \frac{\del}{\del \p_1} + B_2
\frac{\del}{\del \p_2},
\ee

\noindent which in practice involves introducing coordinates
\be
\td{\p}_i = \p_i - B_i z, \qquad \td{\p}_i \equiv \td{\p}_i + 2\pi m_i,
\label{eqn:shifted_coords}
\ee

\noindent  with standard periodicity and which are constant along
orbits of $K$, $K(\td{\p}) = 0$.  In these coordinates, the Killing
vector is simply $K = \del/\del z$.

To show that this identification need not break
supersymmetry~\cite{gutperle:01,farrill:0110}, we work with the shifted coordinates
$\td{\p}$ in the obvious orthonormal basis.  Then it is easy to see
that Killing spinors must have the form
\be
\ep ( \td{\p}_i, z ) =  \exp \left( \frac{1}{2} ( \td{\p}_1
~\Ga_1 + \td{\p}_2 ~\Ga_2 ) \right) \exp \left(
-\frac{1}{2} ( B_1 \Ga_1 + B_2 \Ga_2 ) z \right) \xi_0,
\ee

\noindent $\Ga_i$ being the element of the Lie algebra of ${\rm
Spin}(1,10)$ which generates rotations along $\td{\p_i}$ and has
$\Ga_i^2 = -1$ and where $\xi_0$ is an arbitrary constant spinor.  We
can thus preserve one--half of the $D$--dimensional
supersymmetries if and only if $B_2 = \pm B_1$, the identifications
(\ref{eqn:twotwists}) preserving those spinors which satisfy the
projection condition
\be
\left( 1 \mp \Ga_1 \Ga_2 \right) \xi_0 = 0.
\ee 

Reducing along orbits of $K$, the $(D-1)$--dimensional solution describing
the two--element intersecting fluxbrane is, in the Einstein
frame~\cite{dowker:9512},
\ba
\d s^2_{D-1} &=& \La^{\frac{1}{D-3}} \left( \d s^2 (\mathbb{M}^{D-5}) + \d \rho_1^2 +\ d
\rho_2^2 \right) + \La^{-\frac{D-4}{D-3}} \left( \rho_1^2 \d \p_1^2 + \rho_2^2 \d
\p_2^2 + \rho_1^2 \rho_2^2 \left( B_2 \d \p_1 - B_1 \d \p_2 \right)^2
\right), \nonumber \\
A &=& \La^{-1} \left( \rho_1^2
B_1 \d \p_1 + \rho_2^2 B_2 \d \p_2 \right),\\
e^{\frac{4\p}{\sqrt{D-2}}} &=& \La \equiv 1 + B_1^2 \rho_1^2 + B_2^2 \rho_2^2. \nonumber
\ea

\noindent For $D=11$, this has been referred to as an F5-brane in the
literature~\cite{gutperle:01,farrill:0110}, due to the Poincar\'{e}
invariance in a six--dimensional ``worldvolume'', although strictly speaking this
terminology has been applied only to the supersymmetric case;
the polarization of D-branes due to an F5-brane was studied in
\cite{bena:02}.
The
field strength\footnote{In the same way that the single fluxbrane describes a ``constant''
two--form magnetic field in general relativity, one might think that
the above intersecting fluxbrane in some sense describes a ``constant'' four--form
magnetic field and this, of course, is consistent with the
terminology.  However, there is only ever a two--form in the game, so
it is unclear to what extent the intersecting fluxbrane is an independent entity.}

\noindent has non--vanishing second Chern class~\cite{gutperle:01}
\be
\int_{\mathbb{R}^2 \times \mathbb{R}^2} F \wedge F =
\int_{S^3_{\infty}} A \wedge \d A  = \frac{(2\pi)^2}{B_1 B_2}.
\ee

As for the single fluxbrane~\cite{dowker:9507,dowker:9512},
the intersecting solution should be thought of as a good description
of the configuration of magnetic fields only when they are, in some
sense, small.  For the $(D-1)$--dimensional solution to be weakly curved,
and, when $D=11$, for string theory to be weakly coupled, we need both the $\rho_i
<< B_i$.  But for the Kaluza--Klein ansatz to be valid, we need all
length scales to be much larger than the radius of compactification;
that is, $\rho_i >> R$.  Thus, the ten--dimensional solution is valid
only for both $B_i << 1/R$.  For either $B_i \sim 1/R$, and to keep weak
curvature and string coupling, we need $\rho_i >> R$, so the
ten--dimensional description is no longer a good one.

There is an interesting way to view the \emph{supersymmetric}
intersecting fluxbrane~\cite{farrill:0208107,farrill:0208108}, which is worth making
explicit, and shows why it preserves supersymmetry in the first place.
To see this, take flat Minkowski space as in
(\ref{eqn:flat}).  Think of the $\mathbb{R}^4$ spanned by
$\{\rho_i,\p_i\}$ as $\mathbb{R} \times S^3$ and the $S^3$ as an
$U(1)$ bundle over $S^2$.  Then one can show that the Killing vector
\be
K = B \left( \frac{\del}{\del \p_1} + \frac{\del}{\del \p_2} \right),
\ee

\noindent generates translations along the $U(1)$ Hopf
fibre~\cite{dowker:9507,farrill:0208107,farrill:0208108}.  In fact, identifying points along the
closed orbits of $K$ gives the near--core geometry of Taub--NUT
space~\cite{dowker:9507}.  Introduce the coordinates
\be
\psi = \frac{\p_1}{B}, \qquad \p = \p_2 - B \psi,
\ee

\noindent so that $K = \del / \del \psi$.  Then, upon setting
\be
\rho_1 = \frac{1}{B} \sqrt{\frac{r}{\mu}} \cos (\th/2), \qquad \rho_2 =
\frac{1}{B} \sqrt{\frac{r}{\mu}} \sin (\th/2),
\ee

\noindent and taking $B = 1/(2\mu)$, the metric (\ref{eqn:11Dflat})
becomes
\be
\d s^2_D = \d s^2 (\mathbb{M}^{D-5}) + \d z^2 + V\inv \left( \d \psi + \mu (1 - \cos
\th) \d \p \right)^2 + V ( \d r^2 + r^2 \d \Om_2^2 ),
\ee

\noindent where $V = \mu/r$.  This is easily recognized as the
near--core geometry of the $D$--dimensional Kaluza--Klein
monopole~\cite{gross:83,sorkin:83},
with points identified along orbits of $K = \del / \del \psi$
\emph{i.e.} along the Hopf fibre as promised.

To generate the two intersecting fluxbranes, we simply add a translation along $z$ to the
Killing vector $K$, identifying points along orbits of
\be
K' = \frac{\del}{\del z} + K = \frac{\del}{\del z} + \frac{\del}{\del
\psi}.
\ee

\noindent Since the double twist involved in generating the above fluxbrane
intersection is in the Hopf fibre direction $\psi$, supersymmetry is
guaranteed; none of the Killing spinors of (the near--core region of)
Taub--NUT space are broken by the twist.
If we now set $\t = \psi - z$ such that $K'(\t) = 0$, and reduce
along orbits of $K' = \del/ \del z$ as above, we find, in the Einstein
frame
\ba
\d s^2 &=& H^{\frac{1}{D-3}} V^{\frac{D-4}{D-3}} \left( V\inv \d s^2 (\mathbb{M}^{D-5}) + (HV)\inv
\left( \d \t + \om \right)^2 + \d s^2 (\mathbb{E}^3) \right),\nonumber
\\
A &=& H\inv \left( \d \t + \om \right), \\
e^{\frac{4\p}{\sqrt{D-2}}} &=& H V\inv,\nonumber
\ea

\noindent where $H = 1 + V$, $V = \mu/r$, $\om = \mu (1 - \cos \th)\d \p$
and we have again taken $B = 1/(2\mu)$.  As discussed
in~\cite{farrill:0208107,farrill:0208108}, for $D=11$, this form of
the F5-brane is reminiscent
of that describing bound states of D6-branes and Kaluza--Klein monopoles
found in~\cite{costa:96} where the charges, $\mu$, of both elements
are equal.  The function $V$ would be associated with the
D6-brane, and the function $H$ with the monopole, so the
D6-brane is actually wrapped on the monopole circle.  However, the
match cannot be made exact and we
have also gone to the near--core of the D6-brane, but not of the
monopole, despite the fact that the charges are the same.  In fact,
the above form of the F5-brane is what one would find by taking this
``near--core'' limit of the fluxbranes considered in~\cite{uranga:01,farrill:0208108}.
In these papers, supersymmetric fluxbranes on curved space were found \emph{via}
twisted reductions of the Taub--NUT geometry, whereas to generate the
F5-brane itself, one simply starts instead with the near--core limit
of the Taub--NUT space.

%%%%%%%%%%%%%%%%%%%%%%%%%%%%%%%%%%%%%%%%%%%%%%%%%%%%%%%%%%%%%%%%%%%

%%%%%%%%%%%%%%%%%%%%%%%%%%%%%%%%%%%%%%%%%%%%%%%%%%%%%%%%%%%%%%%%%%%%

\sect{Instantons for intersecting fluxbranes}
\label{sec:instanton}

In general, one should expect similar instabilities of the intersecting
fluxbrane solution as for the single fluxbrane.
Indeed, the relevant decay modes and corresponding instantons have already been
briefly described~\cite{dowker:9512}: decay \emph{via} Witten's
bubble of nothing; nucleation of a locally charged $(D-5)$--sphere in either one of
the fluxbrane elements; or nucleation of an uncharged $(D-7)$--sphere in
the intersecting region
\footnote{With $D=11$, one should have a nice
interpretation of this four--sphere in terms of branes, but it is unclear to us as to
what this should be.  One clue is the
reduction of the ten--dimensional Euclidean Schwarzschild solution
with a trivial time direction. In the
case of a single twist, this gives a six--sphere held in metastable
equilibrium by the background flux~\cite{dowker:9507,dowker:9512}.
By going near the core, one can see explicitly that this is a
spherical D6-brane.  We
have looked at the case with two twists instead, in which one can
indeed identify a four--sphere in the region of intersection, but have
been unable to give this a satisfactory interpretation in terms of
branes.  This is related to our discussion of the F5-brane in section
\ref{sec:f5} which is similar to a bound state of D6-branes and
monopoles.}.  
However, the authors of~\cite{dowker:9512}
did not appreciate that, for $B_1 = \pm B_2$,
the intersecting solution is supersymmetric.  In this case, we would
expect stability, and so would not expect to find such instantons
\footnote{Of course, we are concerned here with non--perturbative stability.  
One should also expect perturbative
stability of the supersymmetric intersecting solutions.  However,
they are not asymptotically flat so, as for supersymmetric plane waves~\cite{brecher:0210,marolf:02}, it
is unclear as to what extent the
presence of Killing spinors guarantees stability.}.

Our starting point is the Myers--Perry~\cite{myers:86} black hole in
arbitrary odd dimension $D=1+N$, with angular momenta, $a_i$, in two
orthogonal planes.  Ultimately, we will be interested in taking
$D=11$.  Analytically continuing $t \ra iz$ and $a_i \ra i\al_i$, the
Euclidean metric is
\[
\d s^2_D = \d z^2 + (r^2-\al_1^2) \left( \d \mu_1^2 + \mu_1^2
\d\p_1^2 \right) + (r^2-\al_2^2) \left( \d \mu_2^2 + \mu_2^2
\d\p_2^2 \right) + r^2 \sum_{i=3}^{N/2} \left( \d \mu^2_i + \mu_i^2
\d\p_i^2 \right)
\]
\be
- \frac{\mu r^2}{\Pi F} \left( \d z + \al_1 \mu_1^2
\d \p_1 + \al_2 \mu_2^2 \d \p_2 \right)^2 + \frac{\Pi F}{\Pi - \mu
r^2} \d r^2,
\label{eqn:instanton}
\ee

\noindent where the direction cosines are constrained as
\be
\mu_1^2 + \mu_2^2 + \sum_{i=3}^{N/2} \mu_i^2 = 1,
\label{eqn:constraint}
\ee

\noindent and 
\be
\Pi = r^{N-4} (r^2-\al_1^2) (r^2-\al_2^2), \qquad F = 1 +
\frac{\al_1^2 \mu_1^2}{(r^2-\al_1^2)} + \frac{\al_2^2
\mu_2^2}{(r^2-\al_2^2)}.
\ee

\noindent There is a ``bolt'' at the origin, $r=r_H$, of polar
coordinates, given by the largest root of $\Pi (r_H) = \mu r_H^2$, so that
\be
\mu = r_H^{N-6} (r_H^2 - \al_1^2) (r_H^2 - \al_2^2),
\label{eqn:origin}
\ee

\noindent and where $\mu \ge 0$.  To avoid a conical singularity at
$r=r_H$, we have to identify the coordinates as
\be
z \equiv z + 2 \pi R, \qquad \p_1 \equiv \p_1 + 2 \pi n \Om_1 R + 2
\pi m_1, \qquad \p_2 \equiv \p_2 + 2 \pi n \Om_2 R + 2 \pi m_2,
\label{eqn:identify}
\ee

\noindent where the radius of the $z$ circle is given by
\be
R = \frac{2\mu r_H^2}{\Pi'(r_H) - 2 \mu r_H} = 2 \mu
\frac{1}{r_H^{N-5}} \left( (N-2)r_H^2 - (N-4)(\al_1^2 +
\al_2^2) + (N-6) \frac{\al_1^2 \al_2^2}{r_H^2} \right)\inv,
\label{eqn:radius}
\ee

\noindent and the ``Euclidean angular velocities'' are
\ba
\Om_1 &=& \frac{\alpha_1}{r_H^2-\alpha_1^2} = \frac{\al_1}{\mu}
(r_H^2 - \al_2^2)r_H^{N-6}, \\
\Om_2 &=& \frac{\alpha_2}{r_H^2-\alpha_2^2} = \frac{\al_2}{\mu}
(r_H^2 - \al_1^2)r_H^{N-6}.
\ea

\noindent Asymptotically, the solution tends to flat space, but
with the non--standard identifications (\ref{eqn:identify}).  In other
words, asymptotically the solution looks like
the Euclidean version of the $D$--dimensional intersecting fluxbrane
solution considered above.  We simply identify the magnetic fields as
\be
B_1 = \Om_1, \qquad B_2 = \Om_2,
\ee

\noindent and we have the correct asymptotics for this to be a
valid instanton describing various decay modes of the intersecting fluxbrane.

An important question concerns the parameter space of this instanton.
That is, what are the possible ranges of the $\Om_i$?   To analyse
this, we first consider the two obvious limits:
\be
(i) ~\al_1 \ra \pm \infty, \qquad \al_2 = {\rm constant} << \al_1,
\ee
\be
(ii) ~\al_2 \ra \pm \infty, \qquad \al_1 = {\rm constant} << \al_2. 
\ee

\noindent From (\ref{eqn:origin}), we must have $r_H \ra |\al_1|$ ($r_H
\ra |\al_2|$) in case $(i)$ ($(ii)$).  Then, from (\ref{eqn:radius}),
we can keep the radius fixed if we take $\mu \ra R |\al_2|^{N-3}$ ($\mu \ra R
|\al_1|^{N-3}$) in case $(i)$ ($(ii)$).  This gives
\be
(i) ~ \Om_1 R \ra \pm 1, \qquad \Om_2 \ra 0,
\ee
\be
(ii) ~ \Om_2 R \ra \pm 1, \qquad \Om_1 \ra 0.
\ee

\noindent These are the two limits in which the parameters effectively
reduce to those of the case with a single rotation, with one of the
$\Om_i = 0$.  

To understand what happens in the intermediate cases, the obvious
limit to take is that in which both
$\al_1$ and $\al_2$ go to infinity at the same rate, namely
\be
\al_1 = \al_2 \equiv \al \ra \infty.
\ee

\noindent Then (\ref{eqn:origin}) gives $r_H \ra \al$ as before, but the
radius (\ref{eqn:radius}) would seem to be ill-defined.  To analyse
this case further, and with an eye on the evaluation of the action in
the following section, we introduce the dimensionless parameters
\be
\hat{\al}_i = \frac{\al_i}{r_H}, \qquad  \hat{\Om}_i = \Om_i r_H, \qquad
\hat{\mu}=\frac{\mu}{r_H^{N-2}}, \qquad \hat{R}=\frac{R}{r_H},
\ee

\noindent in terms of which
\ba
\hat{\mu} &=& (1-\hat{\al}_1^2)(1-\hat{\al}_2^2), \qquad \hat{\Omega}_i = \frac{\hat{\al}_i}{1 - \hat{\al}_i^2}, \\
\hat{R} &=& \frac{2\hat{\mu}}{N-2 - (N-4)(\hat{\al}_1^2 + \hat{\al}_2^2) + (N-6)
\hat{\al}_1^2\hat{\al}_2^2},
\label{eqn:mu+R}
\ea

\noindent so that
\be
\Om_1 R = \hat{\Om}_1 \hat{R} = 2 \hat{\al}_1 (1 - \hat{\al}_2^2)
\left( N-2 - (N-4)(\hat{\al}_1^2
+ \hat{\al}_2^2) + (N-6) \hat{\al}_1^2 \hat{\al}_2^2 \right)\inv,
\label{eqn:omega1}
\ee
\be
\Om_2 R = \hat{\Om}_2 \hat{R} = 2 \hat{\al}_2 (1 - \hat{\al}_1^2)
\left( N-2 - (N-4)(\hat{\al}_1^2 + \hat{\al}_2^2) + (N-6)
\hat{\al}_1^2 \hat{\al}_2^2 \right)\inv.
\label{eqn:omega2}
\ee

\noindent Note that for $\mu \ge 0$, we need either both
$|\hat{\al}_i| \le 1$ or both $|\hat{\al}_i| \ge 1$, but for
$\hat{R}$ positive we are restricted to $|\hat{\al}_i| \le 1$.
Of the four possible combinations of limiting cases
($\hat{\al}_1\rightarrow \pm1$, $\hat{\al}_2\rightarrow \pm1$)
we shall consider the limit $\hat{\alpha}_1 \sim \hat{\alpha}_2 \sim 1$,
so take
\be
\hat{\al}_1 = 1 - \ep_1, \qquad \hat{\al}_2 = 1 - \ep_2,
\label{eqn:limit}
\ee

\noindent where $\ep_1,\ep_2$ are small and positive.  We then find that
\be
\Om_1 R = \frac{\ep_2}{\ep_1 + \ep_2}, \qquad \Om_2 R =
\frac{\ep_1}{\ep_1 + \ep_2},
\label{eqn:boundary}
\ee

\noindent This shows that the limit $\hat{\alpha}_1 \sim \hat{\alpha}_2 \sim 1$
corresponds to $\Om_1 R + \Om_2 R = 1$. Taking the other three
possibilities for the $\hat{\alpha}_i$ limits we find that the
$\Omega_i R$ are bounded to lie in the diamond region defined by the
vertices $(\Om_1 R, \Om_2R) = (0,\pm 1),(\pm 1,0)$.

To higher order one finds
that
\ba
\Om_1 R + \Om_2 R = 1 - \ep_1 \ep_2 \qquad(N=4), \label{eqn:higher1} \\
\Om_1 R + \Om_2 R = 1 - (N-4) \frac{\ep_1\ep_2}{\ep_1 + \ep_2}
\qquad(N>4).
\label{eqn:higher2}
\ea

\noindent If we wish to reach the boundary
($\Om_1 R + \Om_2 R = 1$) with both $\Om_1 R$ and $\Om_2 R$ non-zero
then we see from (\ref{eqn:boundary}) that we can take 
$\epsilon_1=\kappa \epsilon_2$ and then take $\epsilon_2\rightarrow 0$.
So, in the sense of a limiting procedure, we may find the instanton
which has parameters on the boundary of the fundamental region
defined by the diamond with vertices 
$(\Om_1 R, \Om_2R) = (0,\pm 1),(\pm 1,0)$; this
will be relevant when we consider the supersymmetric intersections.

%%%%%%%%%%%%%%%%%%%%%%%%%%%%%%%%%%%%%%%%%%%%%%%%%%%%%%%%%%%%%%%

\sect{Spin structures and decay modes}
\label{sec:spin}

At any rate, just as in the case of a single rotation, there are
various possibilities when we come to dimensionally reduce
the instanton~\cite{dowker:9512}.  In each case, demanding that the $(D-1)$--dimensional
solution be free from conical singularities will restrict the magnetic
field.  Since we can shift either of the $\hat{\Om}_i$
by at most $\pm 1$, there are four cases to consider.  Each describes
a different decay mode, one relevant to either the IIA or 0A theory.
To describe the subsequent evolution of the intersecting fluxbranes,
we should analytically continue one of the ignorable angles into time,
$t$.  The fixed point set of the Killing vector along which
we reduce in each case, restricted to the initial $t=0$ surface, will
give rise to a naked singularity, which can be interpreted as the
``worldvolume'' of a brane in the usual way.
\newcounter{reduce}
\begin{list}
{\arabic{reduce}}{\usecounter{reduce}\setlength{\rightmargin}{\leftmargin}}
\item Reduction along
\[
K = \frac{\del}{\del z} + \Omega_1 \frac{\del}{\del \p_1} + \Omega_2
\frac{\del}{\del \p_2},
\]

\noindent will give a magnetic field $B_1=\Om_1, B_2 = \Om_2$ and a
fixed point set corresponding to the entire origin, a
$(D-2)$--sphere on the $t=0$ surface.  This is Witten's bubble of
nothing~\cite{witten:82}.  
\item Reduction along
\[
K' = K \mp \frac{1}{R} \frac{\del}{\del \p_1},
\]

\noindent will give a magnetic field of $B_1 = \Om_1 \mp 1/R$,
$B_2=\Om_2$ and the fixed point set will be a $(D-5)$--sphere
expanding in one
of the fluxbranes (the one associated with $B_1$).  
\item Reduction along
\[
K' = K \mp \frac{1}{R} \frac{\del}{\del \p_2},
\]

\noindent will give a magnetic field of $B_1 = \Om_1$, $B_2=\Om_2\mp
1/R$ and the fixed point set will again be a $(D-5)$--sphere, but now
expanding in the fluxbrane associated with $B_2$.
\item Reduction along
\[
K' = K \mp \frac{1}{R} \frac{\del}{\del \p_1} \mp \frac{1}{R}
\frac{\del}{\del \p_2},
\]

\noindent will give a magnetic field of $B_1 = \Om_1\mp 1/R$, $B_2=\Om_2\mp
1/R$ and the fixed point set will now be a $(D-7)$--sphere expanding in
the region of intersection.
\end{list}

Again, as for the case of a single rotation~\cite{costa:0012}, the
spin structures determine which of these particular decay modes is
allowed for any given solution.  To see this, we compute the
phase which a spinor picks up under parallel transport at infinity around the
orbits of the Killing vector $K$. 
We first define the shifted
angular coordinates analogous to (\ref{eqn:shifted_coords})
\be
\p_1 = \td{\p}_1 + \Om_1 z, \qquad \p_2 = \td{\p}_2 + \Om_2 z,
\ee

\noindent so that $K = \del /\del z$ and then send $r \ra \infty$ in
the metric (\ref{eqn:instanton}).  The covariant derivative is then
\be
D_z = \del_z + \half \Om_1 \Ga_1 + \half \Om_2 \Ga_2.
\ee

\noindent Parallel transport around a closed loop thus gives
\be
\psi(2 \pi R) = - \exp \left( - \pi R ( \Om_1 \Ga_1 + \Om_2 \Ga_2
)\right) \psi_0,
\label{eqn:phase}
\ee

\noindent where, since the instanton is simply connected
(it is topologically $\mathbb{R}^2
\times S^{D-2}$) , there is a unique spin structure; the overall
minus sign can be determined by continuity with
the $D$--dimensional Euclidean Schwarzschild solution.

Now we see how the spin structure determines the allowed decay modes
1--4 above, since under each independent shift of the Killing
vector, the phase picks up an extra overall minus sign.  Thus, decay
mode 1 will just see the phase (\ref{eqn:phase}).  For $D=11$, it corresponds
to the non--supersymmetric spin structure of the 0A theory.  Decay
modes 2 and 3, however, have an extra overall minus sign, so
correspond to the supersymmetric spin structure of the IIA theory.
Finally, the decay mode 4 picks up a further minus sign, and so is
again only allowed in the 0A theory.  We thus arrive at the following
conclusions; the F5-brane in the IIA theory can decay via nucleation
of a spherical D6-brane, which expands into one of the two elements of
the intersection (that with the larger of the magnetic
fields); and the F5-brane in the 0A theory can decay via Witten's
bubble of nothing~\cite{witten:82} or via the nucleation of a
four--sphere in the intersection region.  Note that
the final, perhaps most interesting, decay mode is pertinent to the
non--supersymmetric 0A theory only.

We still have not answered the question as to how the decay of the
supersymmetric F5-brane, with $B_1 = \pm B_2$, is eliminated.  To
show that it is not allowed, we consider the action for
the above instanton.  We want to compute the action for the range of
possible parameters, that is for each of
the above four cases.  We
will be especially interested in what happens when $B_1 = \pm B_2$ for, in
that case, we do not expect any decay modes to be possible at all.

%%%%%%%%%%%%%%%%%%%%%%%%%%%%%%%%%%%%%%%%%%%%%%%%%%%%%%%%%%%%%%%%%%%%%%%%%%

\section{The action}
\label{sec:action}

Since the metric (\ref{eqn:instanton}) is Ricci--flat, the only
contribution to the action is the boundary term
\be
I_D = - \frac{1}{8\pi G_D} \int_{r \ra \infty} \d^N x \sqrt{h} ( K -
K_0 ),
\ee

\noindent where $h$ is the determinant of the metric induced on
surfaces of constant $r$ and $K$ is the extrinsic curvature of this
surface.  $K_0$ is the curvature of a reference background, which in
this case is just flat space, corresponding to $\mu = 0$.

To compute the action, we have to take account of the
constraint (\ref{eqn:constraint}) on the $\mu_i$, so we substitute for
\be
\mu_{N/2}^2 = 1 - \left( \mu_1^2 + \mu_2^2 + \sum_{i=3}^{N/2-1} \mu_i^2
\right),
\ee

\noindent and write the metric as
\[
\left. \d s_D^2 \right|_{r={\rm const}} = \left( 1 - \frac{\mu
r^2}{\Pi F} \right)\d z^2 + \sum_{i,j=1,2} \left( (r^2 - \al_i^2) \mu_i^2
\delta_{ij} - \frac{\mu r^2}{\Pi F} \al_i \al_j \mu_i^2 \mu_j^2
\right) \d \p_i \d \p_j  
\]
\be
- 2 \frac{\mu r^2}{\Pi F} \sum_{i,j=1,2}
\al_i \mu_i^2 \d \p_i \d z + \sum_{i,j=1}^{N/2-1} \left( (r^2 - \al_i^2) \delta_{ij} +
r^2\frac{\mu_i \mu_j}{\mu_{N/2}^2} \right) \d \mu_i \d \mu_j + r^2
\sum_{i=3}^{N/2} \mu_i^2 \d\p_i^2,
\ee

\noindent where in the second to last term, we have to remember only
$\al_1$ and $\al_2$ are non--zero.  It is easy to compute the
determinant of the metric in the $\{z,\p_1,\p_2\}$ directions
explicitly, and one can use the identity involving
alternating tensors to compute the determinant in the $\mu_i$
directions.  The result is
\be
h = (\mu_1 \ldots \mu_{N/2-1})^2 \frac{\Pi^2 F}{r^2} \left( 1 -
\frac{\mu r^2}{\Pi} \right).
\ee

\noindent We now proceed as in~\cite{dowker:9507}, using the fact that
$\sqrt{h} K = n(\sqrt{h})$, where $n$ is the unit normal
\be
n = \sqrt{\frac{\Pi - \mu r^2}{\Pi F}} \frac{\del}{\del r},
\ee

\noindent which, with a prime denoting $\del / \del r$, gives
\be
\sqrt{h} K = \frac{1}{2} \frac{\mu_1 \ldots \mu_{N/2-1}}{r^2} \left(
-2 \Pi + r \frac{F'}{F} (\Pi - \mu r^2) + r \frac{\Pi'}{\Pi} (2\Pi -
\mu r^2) \right).
\ee

\noindent Since
\be
\lim_{r \ra \infty} \frac{\sqrt{h}}{\left.\sqrt{h}\right|_{\mu=0}} = 1
- \frac{1}{2} \frac{\mu}{r^{N-2}},
\ee

\noindent we have
\ba
\lim_{r \ra \infty} \sqrt{h} (K-K_0) &=& \lim_{r \ra \infty} \left(
n(\sqrt{h}) - \left. n(\sqrt{h})\right|_{\mu=0} + \frac{1}{2}
\frac{\mu}{r^{N-2}} \left. n(\sqrt{h})\right|_{\mu=0} \right) \nonumber
\\
&=& \lim_{r \ra \infty} \left( \frac{\del}{\del \mu} n(\sqrt{h}) +
\frac{1}{2}\frac{1}{r^{N-2}} \left. n(\sqrt{h})\right|_{\mu=0} \right)
\mu.
\ea

\noindent This gives
\ba
\lim_{r \ra \infty} \sqrt{h} (K-K_0) &=& -\frac{1}{2} \mu_1 \ldots
\mu_{N/2-1} \lim_{r \ra \infty} \left( r \frac{F'}{F} (1 - \frac{1}{2}
\frac{\Pi}{r^N} ) + r \frac{\Pi'}{\Pi} (1 - \frac{\Pi}{r^N}) +
\frac{\Pi}{r^N} \right) \nonumber \\
&=& -\frac{1}{2} \left(\mu_1 \ldots \mu_{N/2-1}\right)\mu.
\ea

\noindent We are ultimately interested in the $(D-1)$--dimensional interpretation as a decay
of magnetic fields so we should consider $G_{D-1} = 2\pi R
G_{D}$ as a constant, giving
\be
I_D = \frac{1}{16 \pi} \frac{\Om_{N-1}}{G_N} \mu,
\ee

\noindent where the volume, $\Om_{N-1}$, of the unit $(N-1)$--sphere
comes from the factor
\be
\Om_{N-1} = \int_0^{2\pi} \d \p_1 \ldots \d \p_{N/2} \int \d \mu_1 \ldots \d
\mu_{N/2-1} \mu_1 \ldots \mu_{N/2-1}.
\ee

\noindent We also want to hold the string coupling
constant fixed so the dimensionless parameter of interest is actually
\be
\hat{I} = \frac{\ka^2_N}{R^{N-2}} \frac{1}{\Om_{N-1}} I =
\frac{1}{2} \frac{\mu}{R^{N-2}} = \frac{1}{2} \frac{\hat{\mu}}{\hat{R}^{N-2}},
\ee

\noindent which is the quantity we will plot.

This is a numerical
problem; we pick values of $(\hat{\Om}_1 \hat{R},\hat{\Om}_2
\hat{R})$ within the region of inequivalent spacetimes,
$-\frac{1}{R} \le B_i < \frac{1}{R}$, then solve the equations
(\ref{eqn:omega1}) and (\ref{eqn:omega2}) for
$(\hat{\al}_1,\hat{\al}_2)$, and finally evaluate the action for these
quantities.  
This numerical approach confirms the earlier argument that solutions
are restricted to lie in a diamond region.
Consider, first, the cases 2 and 3 of section
\ref{sec:spin}, those relevant to the type IIA theory.  The action as
a function of $RB_i$ is in these cases is plotted in Fig. 2 and 3.
Taken together they show that an instanton
exists for every value of $-1<RB_i<1$, \emph{i.e.} all inequivalent fluxbranes
have an instanton solution. Also note that Fig. 2 governs instanton solutions
with $|B_1|>|B_2|$. As such, the relevant instanton for $|B_1|>|B_2|$ is
case 2, \emph{i.e.} the decay of the fluxbrane with magnetic
field $B_1$.
Fig. 5 shows the complete case for IIA,
combining cases
2 and 3: the action for $B_1 = \pm B_2$
diverges.

Similarly, cases 1 and 4, plotted in Fig. 1 and 4, together give the complete picture for
decay in type 0A, shown in Fig. 6.  Each case covers a different part of $RB_i$
parameter space and together they give the complete space, showing
that an instanton exists for every inequivalent value of 
$RB_i$.
The action now diverges along $RB_1=\pm 1 \pm RB_2$, these values being the
type 0A dual of the supersymmetric IIA solutions.
We can thus predict that
there are stable intersecting fluxbranes in the type 0A theory for
these values of the magnetic fields.

To see how the action behaves near the supersymmetric
point, consider the limit
(\ref{eqn:limit}).  In that case, the action is found to be
\be
\frac{\hat{\mu}}{\hat{R}^{N-2}} = 4 \ep_1 \ep_2 \left(\frac{\ep_1 +
\ep_2}{\ep_1 \ep_2} \right)^{N-2},
\ee

\noindent and if we look near a point on the boundary by taking
$\ep_1 = \kappa \ep_2$ we see that
\be
\frac{\hat{\mu}}{\hat{R}^{N-2}} = 4\kappa \left(\frac{1 +
\kappa}{\kappa}\right)^{N-2}\frac{1}{\epsilon_2^{N-4}}.
\ee

\noindent 
So, as the boundary is approached ($\epsilon_2=0$)
the action diverges (if $N>4$)
and the semi--classical decay probability goes to zero.
In the special case of $N=4$ we note that the action has
a finite limit as we approach the boundary,
so how do we square this
with the fact that this situation should be supersymmetric?  As
discussed earlier we expect that the fermion zero modes, which are
present due to supersymmetry, will cause the decay amplitude to vanish.
Exactly what happens in the dual type 0A case is unclear and is
under investigation.
It is not unexpected that the case $N=4$ is singled out, because
the two independent rotation planes use up the whole space. It
is possible therefore that similar effects would occur in higher
dimensions when there are more intersecting fluxbranes.

\begin{figure}
\center
\epsfig{file=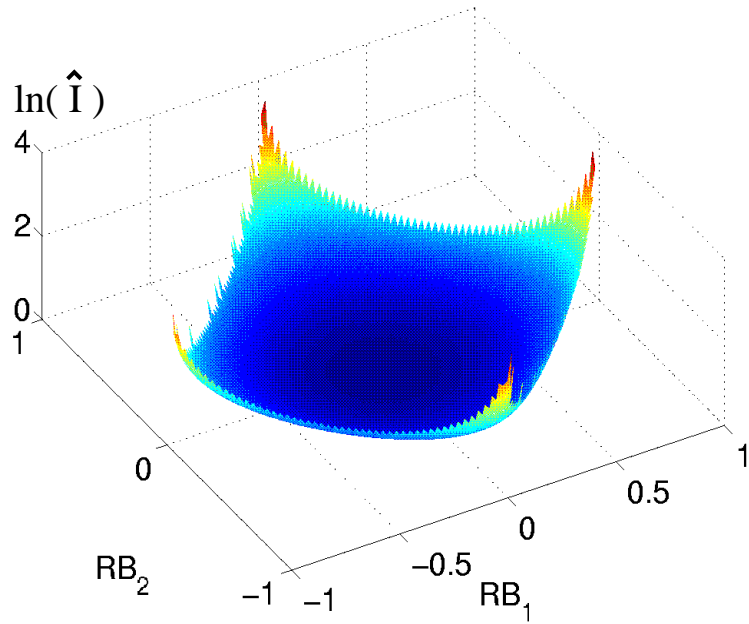,width=10cm}
\flushleft
\caption{This shows the value of the action as a function
         of $RB_i$ for the unshifted instanton (case 1).}
\end{figure}
   
\begin{figure}
\center
\epsfig{file=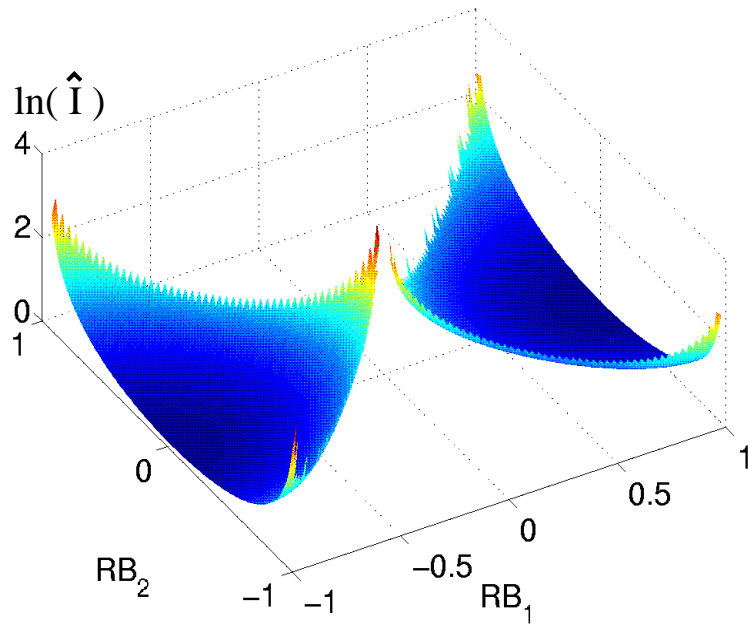,width=10cm}
\flushleft
\caption{This shows the value of the action as a function
         of $RB_i$ for the first shifted instanton (case 2).
         Note that this has $|B_1|>|B_2|$.}
\end{figure}

\begin{figure}
\center
\epsfig{file=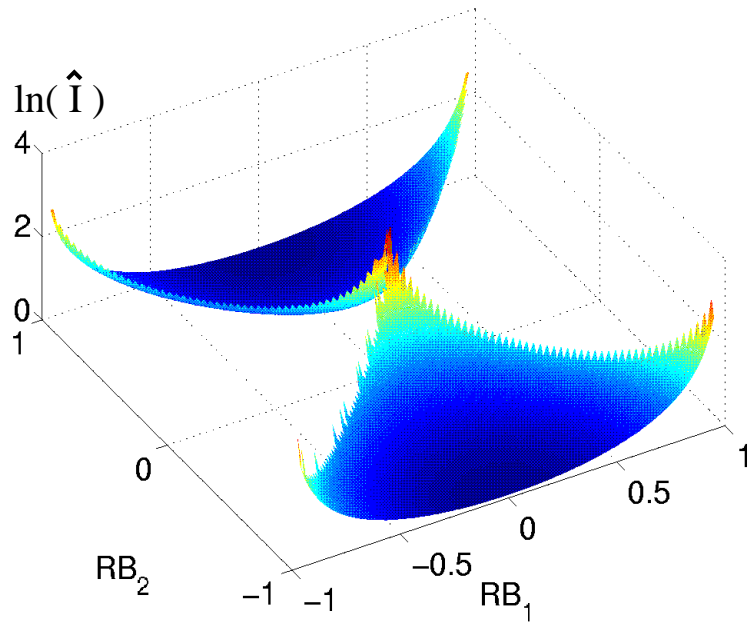,width=10cm}
\flushleft
\caption{This shows the value of the action as a function
         of $RB_i$ for the second shifted instanton (case 3).
         Note that this has $|B_2|>|B_1|$.}
\end{figure}
   
\begin{figure}
\center
\epsfig{file=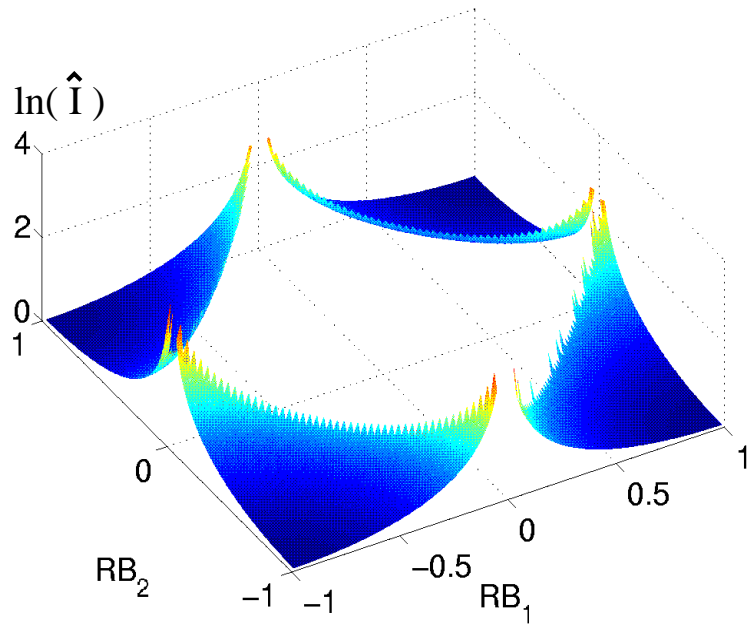,width=10cm}
\flushleft
\caption{This shows the value of the action as a function
         of $RB_i$ for the third shifted instanton (case 4).}
\end{figure}
   
\begin{figure}
\center
\epsfig{file=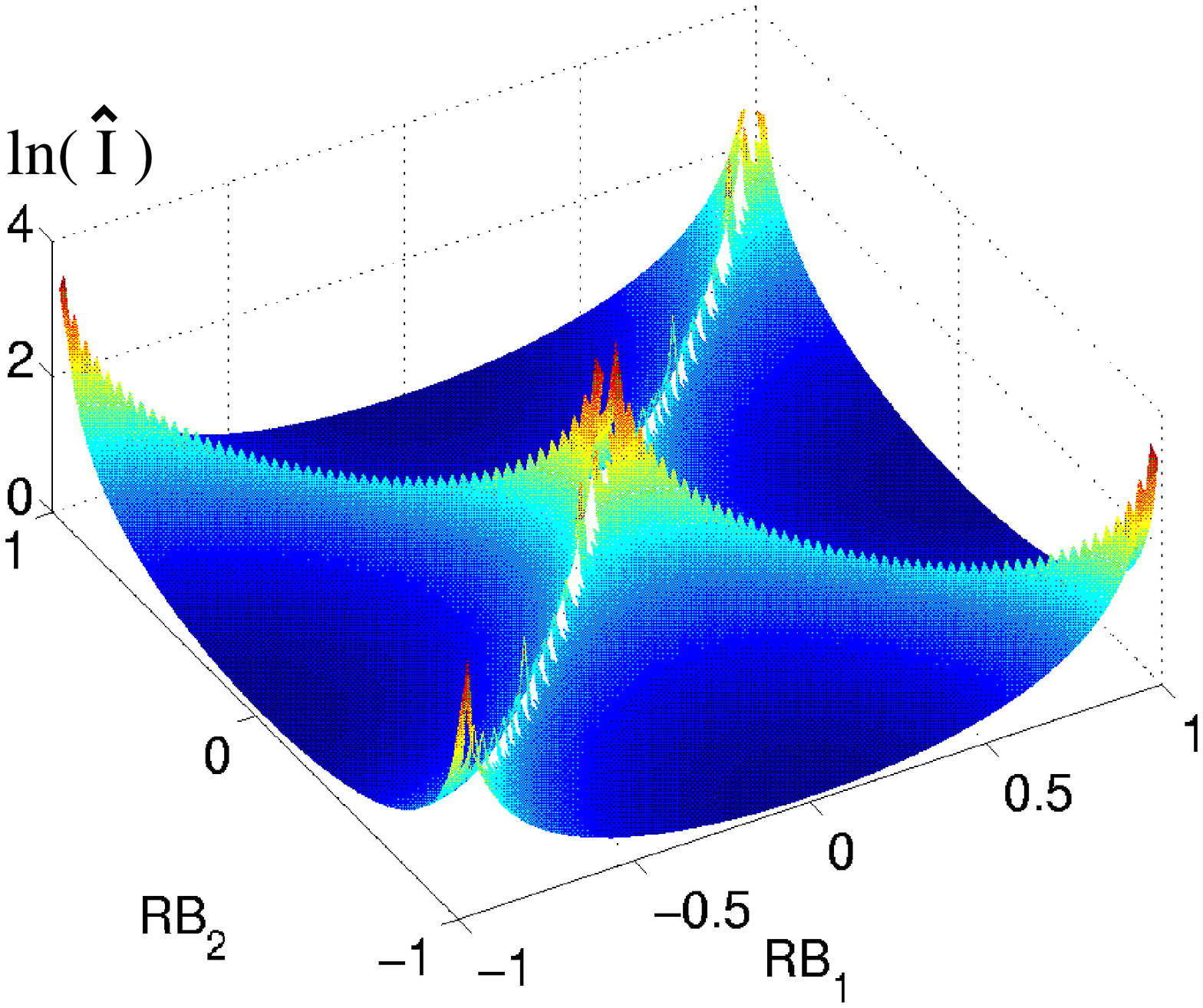,width=10cm}
\flushleft
\caption{This shows the value of the action as a function
         of $RB_i$ for the type IIA theory.}
\end{figure}
   
\begin{figure}
\center
\epsfig{file=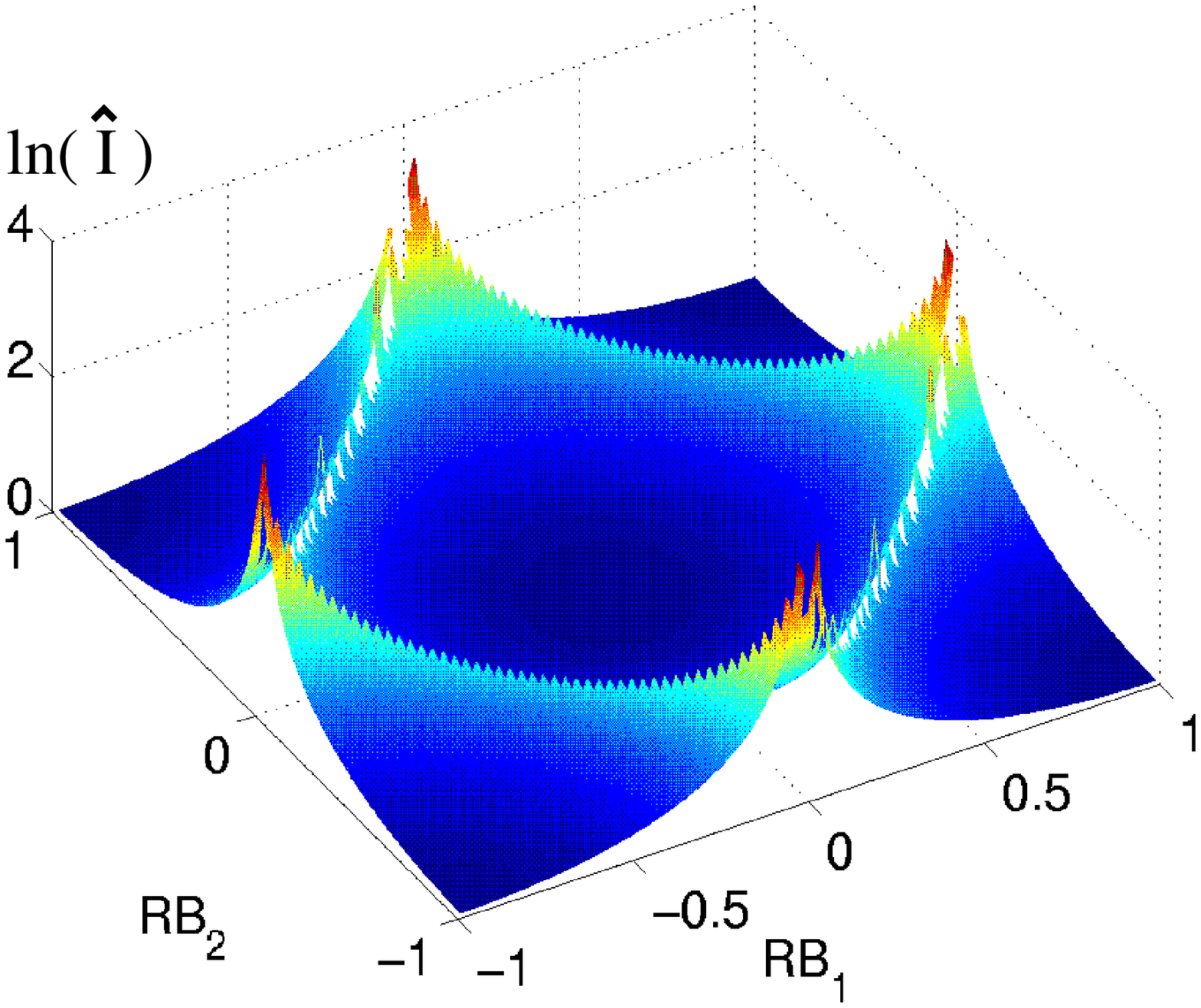,width=10cm}
\flushleft
\caption{This shows the value of the action as a function
         of $RB_i$ for the type 0A theory.}
\end{figure}

%%%%%%%%%%%%%%%%%%%%%%%%%%%%%%%%%%%%%%%%%%%%%%%%%%%%%%%%%%%%%%%%%%%%%%%

\sect{Conclusions}
\label{sec:conclusion}

We have discussed the instantons relevant to the semi--classical decay
of intersecting fluxbranes.  Although we have concentrated on the
two--element intersection, we suspect that similar results hold in
general.  The action, as a function of more than two variables,
however, would soon become difficult to picture.  We have argued that
the parameter space of the instanton includes the
supersymmetric solutions, with $B_1 = \pm B_2$, but that the action
diverges for such solutions.

Of course, due to the presence of fermion zero--modes in the supersymmetric
cases, one can argue that the amplitude for their decay must vanish
identically.  However, it is difficult to verify this without an
explicit computation.  Moreover, by analysing the instanton, we have
been able to argue that stable intersecting fluxbranes in type 0A
string theory (with $B_1=\pm 1/R \pm B_2$) should exist, and arguments about fermion
zero--modes due to supersymmetry would have missed this.  
It would be of interest to think
further about the implications of these stable configurations of the 0A string theory.

It would also be
of interest to see if the interpretation of the F5-brane in terms of
D6-branes and monopoles can be made clearer.  This would give a better
understanding of what it is that is actually nucleated in region of
intersection (decay mode 4 of section \ref{sec:spin}), and might
provide a hints as to a possible ``worldvolume''
theory of the F5-brane.  We have also tried to analyse the late--time
limit of the post--decay evolution of the F5-brane, along the lines
of~\cite{emparan:0111}, but the complicated metric makes it difficult
to extract any interesting results.  Finally, it would be nice to
understand the parameter space of the Euclidean Myers--Perry black
holes for more than two rotation parameters, in relation to
intersecting fluxbranes with more than two elements.

%%%%%%%%%%%%%%%%%%%%%%%%%%%%%%%%%%%%%%%%%%%%%%%%%%%%%%%%%%%%%%%%%%%%%%%%%

\vspace{1cm}
\noindent
{\bf Acknowledgments}\\
We would like to thank Joan Sim\'{o}n for collaboration during the
early stages of this project, and for many useful discussions.
We would also like to thank 
Roberto Emparan, Michael Gutperle and David Mateos for
correspondence.  DB is grateful for the hospitality of the
Perimeter Institute, where much of this work was carried out, and is
supported in part by NSERC.  PMS is supported by PPARC.

%%%%%%%%%%%%%%%%%%%%%%%%%%%%%%%%%%%%%%%%%%%%%%%%%%%%%%%%%%%%

%%%%%%%%%%%%%%%%%%%%%%%%%%%%%%%%%%%%%%%%%%%%%%%%%%%%%%%%%%%%%%%

\end{document}